# A Rural Lens on a Research Agenda for Intelligent Infrastructure

| Ellen Zegura | Beki Grinter | Elizabeth Belding | Klara Nahrstedt |
| --- | --- | --- | --- |
| Georgia Tech | Georgia Tech | University of California, Santa Barbara | University of Illinois at Urbana-Champaign |

A National Agenda for Intelligent Infrastructure is not complete without explicit consideration of the needs of rural communities. While the American population has urbanized, the United States depends on rural communities for agriculture, fishing, forestry, manufacturing and mining. Approximately 20% of the US population lives in rural areas with a skew towards aging adults[1]. Further, nearly 25% of Veterans live in rural America [2]. And yet, when intelligent infrastructure is imagined, it is often done so with implicit or explicit bias towards cities. In this brief we describe the unique opportunities for rural communities and offer an inclusive vision of intelligent infrastructure research.

In this paper, we argue for a set of coordinated actions to ensure that rural Americans are not left behind in this digital revolution. These technological platforms and applications, supported by appropriate policy, will address key issues in transportation, energy, agriculture, public safety and health. We believe that rather than being a set of needs, the rural United States presents a set of exciting possibilities for novel innovation benefiting not just those living there, but the American economy more broadly.

**1.0 The Case for a Rural-Inclusive Intelligent Infrastructure**

Rural communities differ from urban communities in ways that are relevant to intelligent infrastructure considerations. Most obviously, rural communities have sparse population density in comparison to cities. Rather than explicitly defining "rural", the US Census Bureau defines two types of urban areas, those with more than 50,000 people (urbanized areas) and those with 2,500-50,000 people (urban clusters). All other land and populations are considered rural. Under this definition, about 21% of the US population in 2000 was considered rural but more than 95% of the land area was classified as rural. In the 2010 Census, 59.5 million people, 19.3% of the population, was rural while more than 95% of the land area was still classified as rural.[3]

Sparse population densities drive many of the challenges facing rural areas; these are problems that differ from high-density urban areas. They often lack the range of services that a city can provide to residents, such as robust public transit, and diversity of options, such as choices for healthcare. Further, for many years most rural communities have seen a decline in employment opportunities, and while that trend has slowed recently, the employment options are not the same as found in cities.[1] That said some rural communities are seeing improving employment opportunities as the recreation industry grows, and it would be a mistake to take a deficit-only view of rural communities. Instead they are the places some people live and others visit; crucially, Americans are dependent on their success (whether it be from relying on agricultural production, or ensuring that aging Americans and veterans have access to healthcare). Further, for an intelligent infrastructure, rural areas also challenge the dominant modes of thinking about the future of what it means to be smart: inviting consideration of different traffic patterns, pollution causes and locations, agricultural monitoring, aging in place, support for veterans, and so forth.

---

[1] USDA (2016) "Rural America at A Glance: 2016 Edition" Economic Information Bulletin, Economic Research Service.
[2] US Census Bureau (2017) "Nearly One-Quarter of Veterans Live in Rural Areas", Census Bureau Reports January 25, 2017. Release Number: CB17-15. https://www.census.gov/newsroom/press-releases/2017/cb17-15.html
[3] US HRSA (2017) "Defining Rural Population", Federal Office of Rural Health Policy. January 2017. https://www.hrsa.gov/ruralhealth/aboutus/definition.html

**2.0 Foundations of Intelligent Rural Infrastructure**

This whitepaper series defines intelligent infrastructure as "the deep embedding of sensing, computing, and communication capabilities into traditional urban and rural physical infrastructures such as roads, buildings, and bridges for the purpose of increasing efficiency, resiliency, and safety."[4] This definition makes clear that the foundation for intelligent infrastructure is the capability to move data from where it is gathered (sensing) to where it is processed (computing) to where it is used, via communication networks. While these networks need not (and perhaps will not) be exclusively Internet-based, Internet access forms the foundation and backbone for intelligent infrastructure, allowing for real-time access and control, as well as efficient methods for gathering and integrating data into decision-making.

The contrast between Internet access rates in rural versus urban areas remains stark. In 2015, the FCC changed the definition of broadband as access to download speeds of 25Mbps and upload speeds of 4Mbps. By this definition, 10% of all Americans lack access to broadband. However, this disparity breaks down into 39% of rural America as compared to only 4% of residents of urban areas.[5] These disparities worsen as you consider the most rural of settings, such as Native American reservations where only 15% of residents regularly use the Internet, primarily due to lack of access. The lack of broadband access in rural spaces is caused by a convergence of factors that are exemplary when we consider challenges for intelligent infrastructure, including remote and difficult terrain, sparse population densities, lack of IT education, the lack of private investment due to concerns about profitability, and inadequate/outdated infrastructure (e.g., old copper-line networks that have fallen into disrepair, and that cannot support today's needed broadband data speeds).

While some of these rural challenges could be solved by upgrading existing technologies (e.g., replacing old copper-line networks with fiber); rurality and all of its conditions could also be a platform for innovation. Designing rural intelligent infrastructure that is robust in the face of power outages, works in the heat of the Mojave and the wetlands of the Atchafalaya swamp, and is located 100s of miles from any form of significant data processing capability or IT experts, presents important socio-technological challenges. Further, as the history of innovation in computing finds, technologies designed with these goals in mind are likely to find other relevant uses, and generate new industry and market opportunities fueling the American economy.[6]

This innovation begins with government investment. Rural communities today experience the same disadvantage with Internet access as they did with electricity access. By the 1930s, over 90% of urban dwellers had electricity, but only 10% of rural Americans did.[7] Companies argued that it was too expensive to bring the grid to these rural areas—much as the private sector argues that it is not profitable to bring today's networking technologies to sparsely populated areas. Rural residents benefited from the Rural Electrification Act (REA) that fueled the development of infrastructure, which in turn encouraged the private sector to invest.[7] Government investment in a rural Internet-based intelligent infrastructure would position American corporations to partner in the development of applications and services, providing rural Americans with the same opportunities as their counterparts in the cities.

---

[4] Mynatt et al. (2017) "A National Research Agenda for Intelligent Infrastructure" CCC Led Whitepapers http://cra.org/ccc/resources/ccc-led-whitepapers/, last accessed April 12, 2017.
[5] West et al. (2016) "Rural and urban America divided by broadband access" TechTank by Brookings. https://www.brookings.edu/blog/techtank/2016/07/18/rural-and-urban-america-divided-by-broadband-access/. Last accessed April 12, 2017.
[6] NAS (2016) "Continuing Innovation in Information Technology: Workshop Report" National Academies of Sciences, Engineering, and Medicine. https://www.nap.edu/catalog/23393/continuing-innovation-in-information-technology-workshop-report, last accessed April 14, 2017.
[7] FERI (2016) "TVA: Electricity for All" http://newdeal.feri.org/tva/tva10.htm Franklin and Eleanor Roosevelt Institute, New Deal Network, last accessed April 10, 2017.

**3.0 The Potential of Intelligent Rural Infrastructure**

Alongside the foundations for intelligent rural infrastructure as described in Section 2, we revisit problem domains that are candidates for increased efficiency, resilience and safety through research-informed development of intelligent infrastructure. We follow Mynatt et al. (2017), similarly organizing opportunities around seven large domains represented in the rows of Table 1 and four levels of integration into municipal decision making represented in the columns of Table 1. For two row categories—intelligent disaster response and intelligent agriculture—the analysis in Mynatt carries over well to the rural setting; we repeat those rows for completeness. Notably, each row develops a focus within the given problem domain to demonstrate how the opportunities change as the use of intelligence moves from descriptive through proactive. The rural focus for each domain is included in the row label.

|  | **Descriptive** | **Prescriptive** | **Predictive** | **Proactive** |
| --- | --- | --- | --- | --- |
| **Intelligent Transportation (Focus: regional airports)** | Monitor regional airport use and passenger travel patterns | Recommend choice in regional airports and drive/fly combinations based on near and real-time weather, road conditions | Anticipate re-routing of flights to regional airports as necessitated by storms; adjust regional capacity on-demand to strengthen the national air system | Compare and plan for most cost effective regional transportation plans; including high speed trains. |
| **Intelligent Energy Management (Focus: alternative energy production)** | Measure energy production and efficiency delivered from rural regions | Monitor nationwide and regional energy demand and adjust delivery based on market conditions | Anticipate mid-term demand and grow or shift production capacity | Develop new solar and wind farms with delivery channels for economic development |
| **Intelligent Public Safety and Security (Focus: maximizing resource use)** | Combine citizen monitoring with remote sensor monitoring for real-time safety dashboard | Adjust public safety officer deployments based on real-time information | Anticipate vulnerable settings; increase monitoring, pre-deploy assets | Coordinate near and long-term plans across communities for full resource sharing, response to peak events |
| **Intelligent Disaster Response (from [1], Focus: floods)** | Real time water levels in flood prone areas | Timely levee management and evacuations as needed | Anticipate flood inundation with low-cost digital terrain maps | Inform National Flood Insurance Program; Inform vulnerable populations |
| **Intelligent City (and Community) Systems (Focus: drinking water safety and public health)** | Map drinking water sources and delivery lines; digitize historical information regarding underground systems | Advise on targeted, cost-effective plans to replace water system components | Combine technology monitoring, citizen reports, and public health indicators to anticipate and respond to water quality crises before they occur | Inform public health and water safety policy to ensure water system quality and security |
| **Intelligent Agriculture (from [1], Focus: seed and farm mgmt)** | Characterize spatial and temporal variability in soil, crop, and weather | Advise based on environmental stressors and crop traits | Forecast crop yield; Anticipate seasonal water needs | Customize management practices and seed selection to local conditions |
| **Intelligent Health (Focus: epidemic opioid use)** | Track opioid use and response at community level; identify geographic hot spots | Advise on public health response based on best practices from similar community settings | Anticipate individual and community-based conditions predictive of increases in addiction; intervene via policing and community support | Model opioid eco-system from original supply through individual use and plan cost-effective locations to intervene |

The table presents an overview of the opportunities presented by a rurally focused intelligent infrastructure. In the remainder of this section we highlight a few more details of the uniqueness of a rural focus on these various areas of infrastructure development.

*Intelligent Public Safety and Security*: real-time monitoring and analysis is more challenging in rural settings. First, the sheer number of monitoring sensors needed to cover the terrain may well prohibit a naïve approach to blanket the region. Further, sensors for rural areas must function in potentially more hostile environments (deserts, forests) than sensors in urban areas. How we collect data from rural sensors, as well as move it to a computation facilities (whether local or in the Internet) for real-time analysis relies on protocols and technologies that can span distances, and are robust to power outages (which may take longer to resolve when they are in sparsely populated areas). To the extent that city-focused solutions may use people as part of the solution (e.g., police officers are part of an urban network that senses crimes) this may need to be rethought in areas where human resources are harder to find. Applications that partner humans and technologies may have to reconsider the balance of automation and human intervention in rural settings.

*Intelligent Disaster Response*: natural disasters can happen in urban areas, but some types of natural disasters are particularly likely in rural areas. Dams breaking impact rural communities immediately, wildfires start in forested lands, dust storms are associated with farmed areas, tornado alley cuts through largely rural communities, and so forth. As citizens become increasingly engaged with disasters—reporting locations and directions—so we can leverage the human-intelligence in rural communities. However, those same people also need help, roads may be damaged, hospitals are fewer and further between, so we need an intelligent infrastructure that can support their needs, while leveraging the opportunities that they provide to officials trying to determine how best to tackle a crisis.

*Intelligent City (and Community) Systems*: small towns and villages comprise communities. Taking community as a central design feature, many of the concepts proposed for cities might have relevance to more rural communities. For example, cities are interested in technologies that identify and report things that need to be fixed (trash bins to be emptied, potholes to be repaired, water safety to be monitored). Why should this be any different in rural communities—indeed, given the more sparse nature of these areas, reporting potential problems early (e.g., cracks in roads/walls) maybe of more importance given that they pose risk but fewer people see them.

**4.0 Next Steps to Rural Intelligent Infrastructure**

Rural America is not a set of needs, but a set of possibilities. Yes, rural communities need and deserve the same applications that urban Americans will have. However, the infrastructure that provides them will need to address problems not present in urban settings. It is critical to experiment with infrastructure that works in harsh conditions, where power is sparse and potentially intermittent, with novel distributed architectures. These investigations require basic computing research, resulting in innovations generalizable to other resource limited environments ranging from disaster response to space exploration. Additionally, with intelligent infrastructure in place, we see significant opportunities for workforce training, and developing new cyber-related industries that employ rural Americans.

**Internet Access: Novel Networking Innovations**

To bring Internet access to rural areas, a number of wireless technologies have been trialed, and have met with varying levels of success. The technical challenges involved in the deployment of a wireless network in rural areas include network planning, protocols, power management, and hardware failures to inaccessibility of network access locations and remote fault diagnosis. Long distance WiFi-based networks, typically utilizing directional antennas, have been used to connect remote communities both within the United States and abroad. More recently, wireless networks operating in white space spectrum (that is, unused TV spectrum) have shown promise. One of the first such White Space broadband deployments occurred in Claudville, VA, a small community of 20,000 people in the Blue Ridge Mountain terrain.[8] Since this 2009 deployment, white space networks have seen operation in a variety of remote contexts. However, many of these networks are still being run as trials, and more research is needed before this technology is ready for widespread utilization. Beyond white space networking as a technology, rural areas are likely to see a collection of other ad-hoc networking and diverse link level technologies combined to provide the best possible connectivity for reasonable capital and recurring costs.

**Distributed and Localized Computing: Local Clouds**

To create rapid-response capabilities and resilience, rural areas will rely on a computation hierarchy that begins with highly localized computing such as provisioned or on-demand edge clouds (even at the personal or home scale) and includes unlimited Internet-based cloud computation and storage as the top level of the hierarchy. Just as in the rural electrification case, where generators are used for resilience, local clouds will play a similar role. Unlike generators, local clouds will see regular everyday use due to their proximity and thus low latency. Algorithms for local and efficient computation will work in concert with heavy-weight computation in the Internet-based cloud for cases where more precision and/or more detailed data analytics are needed.

---

[8] GT (2010) "Unused Television Spectrum Could Deliver New Broadband Services" Government Technology. http://www.govtech.com/templates/gov_print_article?id=99262679, last accessed April 14, 2017.

**Rural Testbeds: Engaging Rural Americans**

The rural United States has a staggering range of difficult terrain to support innovation in intelligent infrastructure. Deserts, swamps, young and old mountain ranges with the variance in topography, prairies, tropical and temperate forests, rainforests, the list is endless. Americans live and visit all these places; they are the rural living laboratories. We need testbeds that will initially challenge our scientific and technological imaginations as we design solutions that can function in these environments and across these vast terrains, as well as the supporting systems that allow local residents to work with them. Testbeds would also provide the platform to explore remote operation and maintenance of infrastructure (e.g., upgrading devices), as well as the advantages and disadvantages of centralized and distributed architectures.

**Education and Workforce Development: Employment Opportunities**

Testbeds serve a second importance, as location for training the local population. The government already has a number of initiatives to retrain American workers for the digital economy, and the demand for digital skills will only grow as intelligent infrastructure evolves. Americans need to be more than just consumers of a smart digital grid, they should be able to maintain and evolve it. In rural communities this is even more important since it is the local population who may end up managing far more than just their own consumer needs. Agricultural, forestry, fishery, and mine workers may play a vital role in maintaining networks of sensors in remote places that serve Americans in rural and urban populations. Indeed, we posit a digital dimension to many of these traditional employment sectors for rural Americans, digital farming may mean the management of watering, crop harvesting and so forth informed by intelligent infrastructure. Thus rural intelligent infrastructure, which many Americans will increasingly rely on, can be the source of education and employment opportunities for rural residents.

Beyond testbeds, rural communities could leverage intelligent infrastructure for education and workforce needs. For example, the recent explosion in digital education (e.g., Massive Online Open Courses), could provide valuable educational opportunities for Americans in rural settings, but only if Internet access is available and sufficient to make theses on-line training opportunities high quality.

**Conclusion**

A National Agenda for Intelligent Infrastructure is not complete without explicit consideration of the needs of rural communities. And yet, Intelligent Infrastructure is often imagined as "smart cities" with bias towards urban needs. We propose a Rural-focused Intelligent Infrastructure Act: technological platforms, applications, supported by appropriate policy that empowers rural communities. At its core, we believe that the rural United States presents exciting possibility for groundbreaking innovations with outcomes that will benefit not just those who live and work there, but the American economy more broadly.

*This material is based upon work supported by the National Science Foundation under Grant No. 1136993. Any opinions, findings, and conclusions or recommendations expressed in this material are those of the authors and do not necessarily reflect the views of the National Science Foundation.*